\documentclass{optica-article}

\journal{opticajournal}
\articletype{Research Article}

\usepackage{amsmath}
\usepackage{xcolor}
\usepackage{lineno}
\usepackage{algorithm}
\usepackage{algorithmicx}
\usepackage{algpseudocode}
\graphicspath{{./images/}}
\let\mathscr\mathcal

\begin{document}

\title{Feature-domain Fourier ptychographic tomography with dark-field illumination}

\author{Chao Tan,\authormark{1}
Fangrui Lu,\authormark{1}
Sechan Park,\authormark{1}
Hyeonseo Na,\authormark{1}
Chanseok Lee,\authormark{1}
Chang-Seok Kim,\authormark{2,3}
Jeesu Kim,\authormark{2,3}
Hwidon Lee,\authormark{2,3}
and Mooseok Jang\authormark{1,4,*}}

\address{\authormark{1}Department of Bio and Brain Engineering, Korea Advanced Institute of Science and Technology (KAIST), Daejeon 34141, Republic of Korea\\
\authormark{2}Department of Cogno-Mechatronics Engineering, Pusan National University, Busan 46241, Republic of Korea\\
\authormark{3}Engineering Research Center for Color-Modulated Extra-Sensory Perception Technology, Pusan National University, Busan 46241, Republic of Korea\\
\authormark{4}KAIST Institute for Health Science and Technology, Korea Advanced Institute of Science and Technology (KAIST), Daejeon 34141, Republic of Korea}

\email{\authormark{*}mooseok@kaist.ac.kr}

\begin{abstract*}
Fourier ptychographic tomography (FPT) is an implementation of intensity diffraction tomography that reconstructs three-dimensional (3D) refractive-index (RI) distributions from angle-varied intensity measurements. The distinctive advantage of FPT emerges when incorporating dark-field illumination, which extends the space-bandwidth product toward gigavoxel-scale volumetric imaging---yet dark-field measurements are highly sensitive to system imperfections and often have low signal-to-noise ratios. Here, we propose feature-domain FPT (FD-FPT), which evaluates data fidelity after feature extraction and is optimized using automatic differentiation. In numerical and experimental tests, FD-FPT resolves structures near the synthetic-aperture cutoff far more reliably than the spatial-domain baseline (SD-FPT). Notably, in a whole-mount \textit{Oedogonium} specimen, the reticulate chloroplast network and transverse septa were resolved only by FD-FPT. We further demonstrate a 1.81-gigavoxel RI reconstruction of a mouse adrenal gland section across a $1.66\times1.40\,\mathrm{mm}^2$ field of view, establishing FD-FPT as a practical route to label-free volumetric imaging that combines millimeter-scale coverage with cellular-scale structural contrast.
\end{abstract*}


\section{Introduction}

With advancements in semiconductors, materials science, and biomedicine, there is an increasing demand for 3D optical imaging techniques that offer high resolution, a large field of view (FOV), and an extended depth of field \cite{hawkes2019springer}. While techniques such as light sheet, confocal, and two-photon microscopy provide high-quality 3D images \cite{mertz2019strategies}, they typically rely on exogenous labeling. This is often inapplicable to solid-state specimens and can also alter the physiological state of biological samples due to phototoxicity and photobleaching \cite{lee2013quantitative}. In contrast, diffraction tomography (DT), first proposed by Wolf in the 1960s \cite{wolf1969three}, utilizes the refractive index (RI) of 3D structures as an intrinsic contrast agent. This label-free approach enables quantitative analysis of cellular characteristics as well as nondestructive metrology and inspection of semiconductor devices.

Generally, the DT reconstruction process consists of two primary stages: retrieving complex fields and reconstructing the 3D RI distribution from these fields \cite{devaney2012mathematical}. Conventional optical diffraction tomography (ODT) relies on interferometric measurements to retrieve complex-field information \cite{lauer2002new,sung2009optical,bhaduri2014diffraction,jin2017tomographic,kamilov2015learning}. These approaches often use coherent sources, high-numerical-aperture (NA) objective lenses, mechanical scanning, and stable experimental setups. Although they can provide high-quality 3D reconstructions, their use can be constrained by a relatively small FOV, coherent artifacts, and system cost \cite{maleki1993phase,gbur2004information,horstmeyer2016diffraction,li2017optical,ling2018high,zuo2020wide,baek2021intensity,li2022transport,zhou2022transport,sun2022high,zhou2020diffraction}. Noninterferometric methods, including the transport-of-intensity equation (TIE) \cite{li2017optical,li2022transport,zhou2022transport}, Kramers--Kronig relations \cite{baek2021intensity,li2022transport}, and Fourier ptychography (FP) \cite{horstmeyer2016diffraction,li2022transport,zhou2022transport,zhou2020diffraction}, have therefore been widely studied. Fourier ptychography can achieve high-resolution, large-FOV imaging using a programmable LED array \cite{zheng2013wide,zheng2021concept,jiang2023spatial,xu2024fourier}. When combined with Fourier diffraction theory (e.g., the Born or Rytov approximations) \cite{wolf1969three} in 3D Fourier space, this ptychographic reconstruction approach is referred to as Fourier ptychographic tomography (FPT) or Fourier ptychographic diffraction tomography (FPDT) and represents an implementation of intensity diffraction tomography (IDT) \cite{horstmeyer2016diffraction,zuo2020wide,sun2022high,zhou2020diffraction}. Other inverse-scattering solvers, such as multislice beam propagation (MSBP) and the modified Born series \cite{tian20153d,li2018multi,chowdhury2019high,lee2022inverse}, have also been used to reconstruct 3D RI distributions.

FPT and Fourier ptychographic microscopy (FPM) differ in the physical models used to describe the interaction between optical fields and the sample. Whereas FPM assumes an infinitesimally thin 2D sample, FPT accounts for 3D inhomogeneity in the RI distribution. The two modalities can nevertheless use the same LED-array hardware, including dark-field illumination at angles beyond the detection numerical aperture (NA). Incorporating dark-field data expands the accessible spatial-frequency spectrum and can therefore improve resolution and reduce the "missing-cone" problem, which results from limited illumination angles and contributes to axial elongation and RI underestimation \cite{lim2015comparative,zuo2020wide,sun2022high,dong2025analytic}. Like FPM, FPT remains susceptible to system imperfections, including LED misalignment, vignetting, and nonuniform illumination \cite{yeh2015experimental,pan2017system,zhang2023elfpie,chen2025uncertainty}. Dark-field measurements can be especially sensitive to these imperfections because adjacent illumination angles have less overlap in 3D Fourier space and the measured signals often have low signal-to-noise ratios (SNRs). Robust reconstruction algorithms or careful data preprocessing are therefore needed to incorporate these measurements reliably.

To address these challenges, we propose feature-domain Fourier ptychographic tomography (FD-FPT). Unlike conventional methods that evaluate data fidelity directly in the spatial domain, FD-FPT evaluates the loss after feature extraction (e.g., with gradient operators) \cite{zhang2025whole,zhang2023elfpie,zhao2024deep,zhang2025high,wu2025wavelet}. This formulation is intended to reduce sensitivity to system-induced intensity variations and to facilitate the incorporation of dark-field data without extensive preprocessing. We implement the reconstruction framework using automatic differentiation (AD) \cite{baydin2018automatic,paszke2017automatic}, which avoids manual gradient derivation and facilitates the integration of regularization terms. We evaluate FD-FPT using simulations of a lymph-node vascular phantom and experimental imaging of a USAF resolution target, polystyrene microspheres, a whole-mount \textit{Oedogonium} specimen, and a large-scale mouse adrenal gland section. The experimental system provides a synthetic NA of approximately 0.63 and an FOV exceeding $1\,\mathrm{mm}\times1\,\mathrm{mm}$, resulting in a $\sim$ 1.81-gigavoxel 3D reconstruction. Comparisons with the spatial-domain baseline assess how the feature-domain loss and the inclusion of dark-field measurements affect resolution, contrast, and missing-cone-related artifacts in label-free 3D imaging \cite{jin2017tomographic,park2018quantitative,nguyen2022quantitative,sun2022high}.

\section{Methods}
\begin{figure*}[!t]
    \centering
    \includegraphics[width=\linewidth]{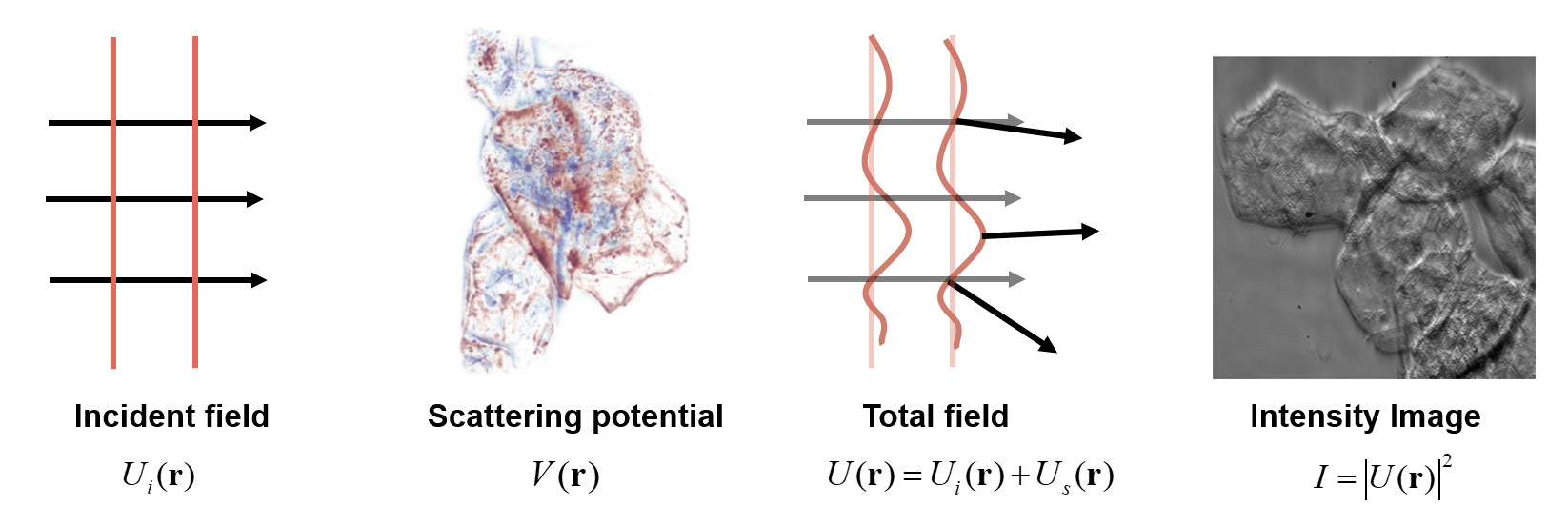}
    \caption{Schematic illustration of the diffraction tomography principle.}
    \label{fig:principle}
\end{figure*}

Figure~\ref{fig:principle} illustrates the diffraction tomography principle. In this section, we briefly introduce the theoretical framework of diffraction tomography (DT) and the principle of our proposed method. We first formulate the interaction between the object and the optical field, based on the inhomogeneous Helmholtz equation for a scalar field:
\begin{equation}
[\nabla^2 + k^2]U(\boldsymbol{r}) = -V(\boldsymbol{r})U(\boldsymbol{r}),
\label{inhomogeneous_Helmholtz}
\end{equation}
where $\nabla^2$ denotes the Laplacian operator, $k_0=2\pi/\lambda$ is the vacuum wavenumber, $k=k_0n_0$ is the wavenumber in the background medium, $U(\boldsymbol{r})$ represents the total field, and $V(\boldsymbol{r})$ is the scattering potential \cite{wolf1969three}. The scattering potential is defined as
$V(\boldsymbol{r}) = k_0^2 \left( n(\boldsymbol{r})^2 - n_0^2 \right)$, where $\boldsymbol{r} = (x,y,z)$ denotes the spatial coordinates, $n(\boldsymbol{r})$ is the 3D refractive-index distribution of the object, and $n_0$ is the refractive index of the surrounding medium. In general, $n(\boldsymbol{r})$ is complex-valued: its real part determines the phase delay, whereas its imaginary part describes optical attenuation. For a monochromatic incident plane wave $U_i(\boldsymbol{r}) = e^{i \boldsymbol{k}_i \cdot \boldsymbol{r}}$, where the incident wave vector satisfies $|\boldsymbol{k}_i|=k$, the total field can be decomposed as $U(\boldsymbol{r})=U_i(\boldsymbol{r}) + U_s(\boldsymbol{r})$, with $U_s(\boldsymbol{r})$ denoting the scattered field. In Eq.~\eqref{inhomogeneous_Helmholtz}, the scattering potential represents the source term arising from the interaction between the object and the wave field \cite{muller2015theory}. The corresponding Lippmann--Schwinger equation is
\begin{equation}
U_s(\boldsymbol{r}) = \int G(\boldsymbol{r} - \boldsymbol{r'}) V(\boldsymbol{r'}) U(\boldsymbol{r'}) \mathrm{d}\boldsymbol{r'},
\label{LS}
\end{equation}
where the Green's function is $G(\boldsymbol{r} - \boldsymbol{r'}) = \frac{e^{ik|\boldsymbol{r} - \boldsymbol{r'}|}}{4\pi|\boldsymbol{r} - \boldsymbol{r'}|}$.

Iteratively substituting $U(\boldsymbol{r})=U_i(\boldsymbol{r}) + U_s(\boldsymbol{r})$ into Eq.~\eqref{LS} yields the Born series. Because evaluating this series is computationally demanding, we use the first Born approximation, which assumes single scattering and is most appropriate when $|U_s(\boldsymbol{r})| \ll |U_i(\boldsymbol{r})|$. Under this condition, $U(\boldsymbol{r}) \approx U_i(\boldsymbol{r})$ on the right-hand side of Eq.~\eqref{LS}, and the scattered field can be expressed using the Fourier diffraction theorem \cite{kak2001principles}:
\begin{equation}
U_s^{(B)}(\boldsymbol{r}) = \mathscr{F}^{-1}\{ \tilde{G}(\boldsymbol{k}) \tilde{V}(\boldsymbol{k} - \boldsymbol{k}_i) \},
\label{FDT}
\end{equation}
where $U_s^{(B)}(\boldsymbol{r})$ is the scattered field under the first Born approximation, $\mathscr{F}^{-1}$ is the inverse Fourier transform, $\tilde{G}(\boldsymbol{k})$ is the Fourier transform of the Green's function, and $\tilde{V}(\boldsymbol{k})$ is the Fourier transform of $V(\boldsymbol{r})$. Equation~\eqref{FDT} shows that varying the illumination angle translates the accessible region of the scattering-potential spectrum by the incident wave vector, allowing multiple angles to sample a wider region of 3D Fourier space.

FPT commonly uses a programmable LED array to provide sequential illumination from different angles. For the $n$-th LED illumination, the ideal forward model is

\begin{equation}
    \hat{I}^n = \left| \mathcal{F}^{-1}\{\mathcal{F}\{U^n(\boldsymbol{r}_\perp)\} \times P\} \right|^2.
    \label{forward}
\end{equation}

Here, $\hat{I}^n$ denotes the ideal 2D intensity image for the $n$-th LED illumination, and $U^n(\boldsymbol{r}_\perp)$ is the total field at the objective focal plane. The vector $\boldsymbol{r}_\perp$ denotes the 2D transverse position, $\mathcal{F}$ denotes the Fourier transform, and $P$ is the pupil function. Although FPT and FPM can use the same experimental setup and similar measurement operators, their object models differ. Under the thin-sample approximation used in FPM, the interaction is represented by pointwise multiplication: $U^n(\boldsymbol{r}_\perp) = U_i^n(\boldsymbol{r}_\perp) O(\boldsymbol{r}_\perp)$, where $O(\boldsymbol{r}_\perp)$ is the 2D object function. In FPT, the Fourier diffraction theorem describes the 3D interaction between the object's RI distribution and the optical field. Expressing Eq.~\eqref{FDT} at the detector plane, the scattered component of $U^n(\boldsymbol{r}_\perp)$ can be written as

\begin{equation}
U_s^{n,(B)}(\boldsymbol{r}_{\perp})
= \mathcal{F}_{2\mathrm{D}}^{-1}\!\left\{
\frac{i}{2k_z}\,
\tilde{V}\!\left(
\boldsymbol{k}_{\perp}-\boldsymbol{k}_{i\perp}^n,
k_z-k_{iz}^n
\right)\right\},
\label{ewald_sampling}
\end{equation}

where $\boldsymbol{k}_i^n = (\boldsymbol{k}_{i\perp}^n, k_{iz}^n)$ denotes the incident wave vector for the $n$-th LED illumination, $\boldsymbol{k}_\perp$ represents the transverse spatial frequencies collected at the pupil plane, and $k_z = \sqrt{k^2 - |\boldsymbol{k}_\perp|^2}$ is the corresponding axial spatial frequency constrained by the Ewald sphere.

\begin{figure*}[!t]
    \centering
    \includegraphics[width=\linewidth]{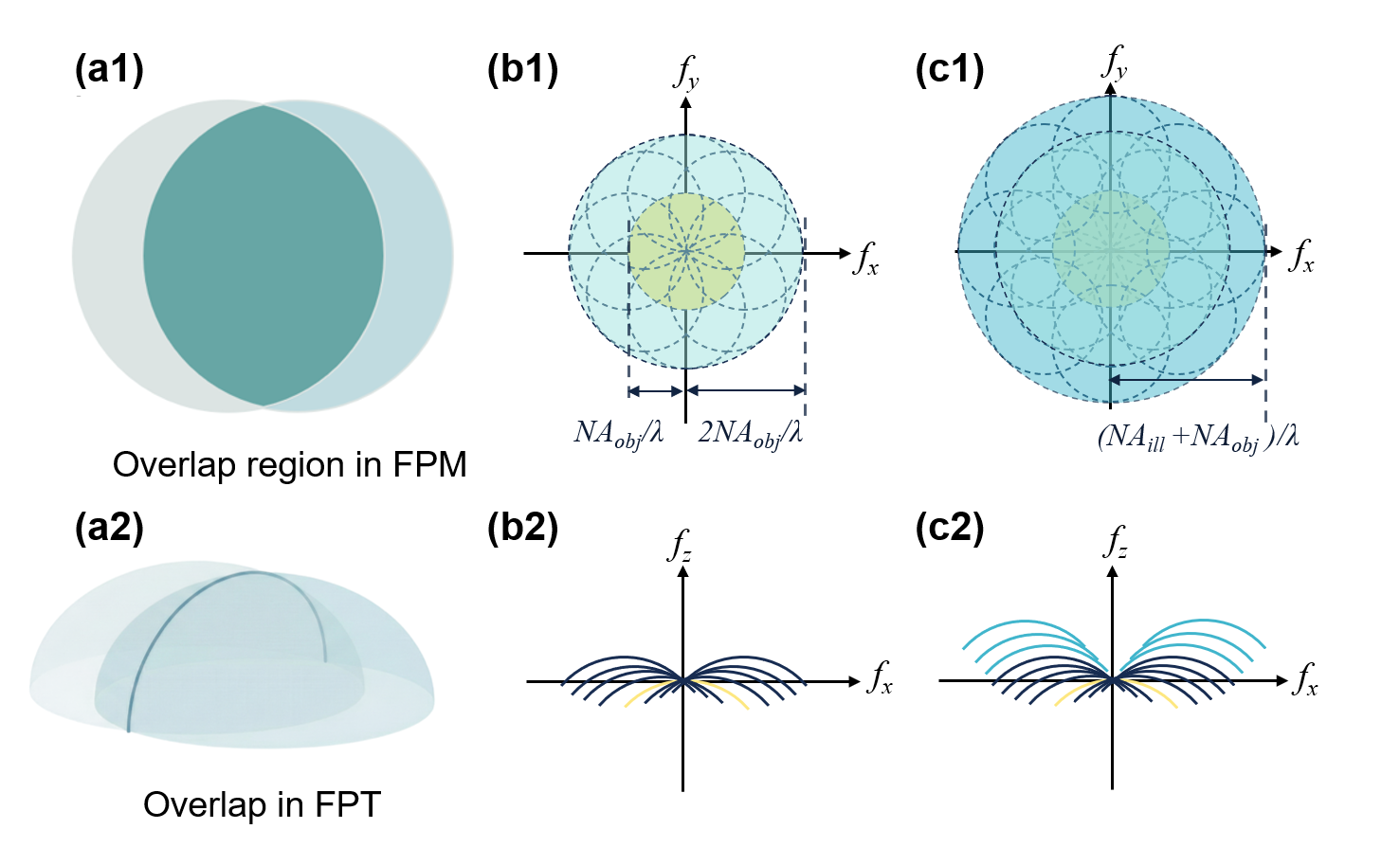}
    \caption{Comparison of spectral overlap and coverage. \textbf{(a)} Spectral-overlap regions for (a1) FPM and (a2) FPT. \textbf{(b)} Limited spectral coverage in the lateral (b1) and axial (b2) planes under bright-field illumination. \textbf{(c)} Extended lateral (c1) and axial (c2) coverage under dark-field illumination.}
    \label{fig:spectral_coverage}
\end{figure*}

Therefore, complex-field retrieval in FPT must account for the 3D nature of the object and the scattered field. As shown in Figure~\ref{fig:spectral_coverage}, the spectral-overlap geometry in FPM differs from that in FPT. FPM involves overlapping 2D circles, whereas FPT involves intersections of Ewald spheres and consequently has less spectral overlap between adjacent illumination angles. This geometry makes recovery of the 3D scattering potential from intensity measurements more challenging because phase retrieval relies on measurement redundancy. The FPT forward model also treats the incident field differently under bright-field and dark-field illumination. Under bright-field illumination, the detected total field is $U_i^n(\boldsymbol{r}_{\perp}) + U_s^{n,(B)}(\boldsymbol{r}_{\perp})$; under dark-field illumination, the unscattered incident field lies outside the objective pupil, and the detected field is $U_s^{n,(B)}(\boldsymbol{r}_{\perp})$.

Equations~\eqref{forward} and~\eqref{ewald_sampling} describe FPT under ideal conditions. In practice, measured intensities are affected by nonuniform illumination, vignetting, model mismatches such as LED-position errors, pupil aberrations, stray light, and detector noise \cite{yeh2015experimental,pan2017system,zhang2023elfpie,chen2025uncertainty}. The measured images can therefore differ from the predicted intensities even when the reconstructed scattering potential captures the main sample structures. This problem is particularly pronounced for dark-field measurements, which are dominated by weak scattered signals. Consequently, directly minimizing the spatial-domain intensity residual, $\left\|\hat{I}^n-I^n \right\|$, may cause the reconstruction to fit measurement artifacts. Reduced spectral overlap in 3D $k$-space [Figure~\ref{fig:spectral_coverage}(a)] further increases sensitivity to measurement inconsistencies. These factors make dark-field data more difficult to incorporate reliably than bright-field data. Excluding dark-field information, however, reduces the accessible spatial-frequency range and limits the ability of FPT to combine a large FOV with high spatial resolution [Figure~\ref{fig:spectral_coverage}(b,c)].

\begin{figure*}[!t]
    \centering
    \includegraphics[width=\linewidth]{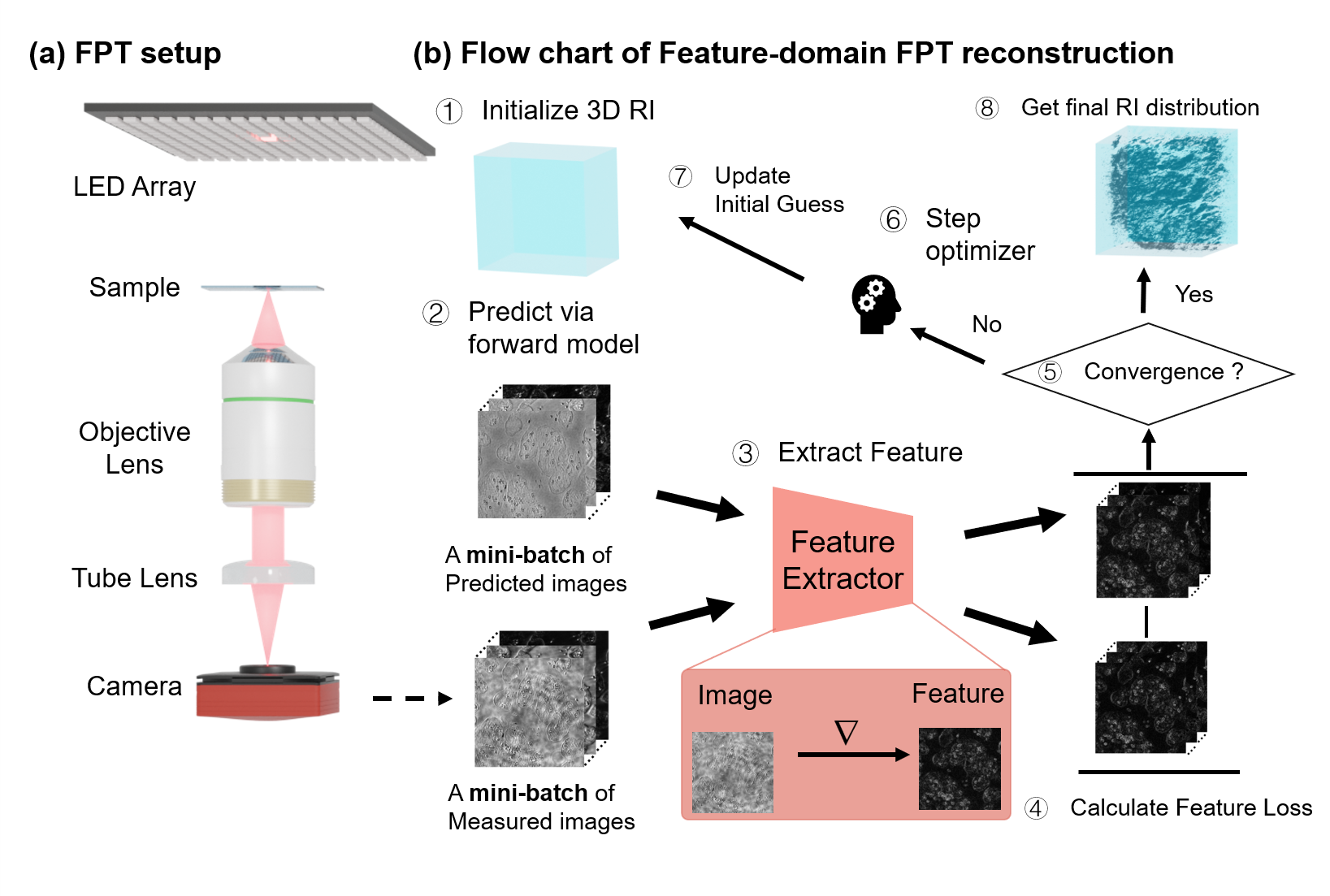}
    \caption{Overview of the FD-FPT framework. (a) Experimental setup using a programmable LED array. (b) Reconstruction flowchart.}
    \label{fig:fd-fpt-pipeline}
\end{figure*}

To address this challenge, we propose feature-domain Fourier ptychographic tomography (FD-FPT), inspired by feature-domain losses used in computational microscopy \cite{zhang2025whole,zhang2023elfpie,zhao2024deep,zhang2025high,wu2025wavelet}. Instead of comparing measured and predicted intensities directly, FD-FPT applies a feature extractor before evaluating their discrepancy. Slowly varying intensity nonuniformity and spatially localized sample structures respond differently to a transverse-gradient operator; the resulting feature-domain loss therefore reduces the relative contribution of low-frequency intensity variations while retaining edge information that can guide the tomographic update (Supplementary Figure~S1). We define the FD-FPT objective as
\begin{equation}
\mathcal{L}_{\text{FD-FPT}}(V)=\sum_{n=1}^{N}\left\|\nabla_{\perp}(I^n)^{\gamma}-\nabla_{\perp}(\hat{I}^n)^{\gamma}\right\|_2^2+\mathscr{R}(V),
\label{obj-fun}
\end{equation}
where $I^n$ and $\hat{I}^n$ are the measured and predicted intensity maps for the $n$-th LED illumination, respectively, and $N$ is the number of illumination angles. The parameter $\gamma$ controls the intensity scaling: $\gamma=0.5$ yields an amplitude-based loss, whereas $\gamma=1$ yields an intensity-based loss. We use the transverse spatial-gradient operator $\nabla_{\perp}$ as a computationally efficient, training-free feature extractor. The regularization term $\mathscr{R}(V)$ comprises the total-variation (TV) and Hessian penalties used in the reconstruction.
Minimizing this objective yields an estimate of the scattering potential ${V}(\boldsymbol{r})$, which is converted to the 3D refractive-index distribution using $n(\boldsymbol{r})=\sqrt{n_0^2+V(\boldsymbol{r})/k_0^2}$. The FD-FPT pipeline is illustrated in Figure~\ref{fig:fd-fpt-pipeline}(b), and Algorithm~\ref{alg:fdfpt_reconstruction} summarizes the reconstruction procedure. We iteratively update the 3D scattering-potential spectrum $\tilde{V}$ using the RMSprop optimizer, with gradients computed by automatic differentiation in PyTorch. All computations reported below were performed on a workstation equipped with an NVIDIA GeForce RTX 4090 GPU (24\,GB VRAM).

\begin{algorithm*}[t]
\caption{FD-FPT reconstruction algorithm}
\label{alg:fdfpt_reconstruction}
\begingroup
\footnotesize
\setlength{\baselineskip}{8.8pt}
\begin{algorithmic}[1]
\Require Measurements $\{\boldsymbol{I}^n\}$, incident fields $\{U_i^n\}$, epochs $E$, batch size $B$, pupil $P(\boldsymbol{k}_{\perp})$, medium RI $n_0$, vacuum wavenumber $k_0$, and background-medium wavenumber $k$
\Ensure Reconstructed refractive-index distribution $n(\boldsymbol{r})$
\State Initialize $\tilde{V}_0(\boldsymbol{k})$ with background values and set $t \leftarrow 0$
\State Precompute $F_{\mathrm{meas}}^n \leftarrow \nabla_{\perp}[(\boldsymbol{I}^n)^{\gamma}]$ for all $n$
\For{$e=1,\ldots,E$}
    \For{each batch of indices $\mathcal{B}$ with size $B$}
        \State $\tilde{U}_s^{\mathcal{B}} \leftarrow \operatorname{SampleKSpace}(\tilde{V}_t,\mathcal{B})$ using Eq.~\eqref{ewald_sampling}
        \State $U_{\mathrm{filt}} \leftarrow \mathcal{F}_{2\mathrm{D}}^{-1}\{\tilde{U}_s^{\mathcal{B}}P(\boldsymbol{k}_{\perp})\}$
        \State Retrieve the incident fields $U_i^{\mathcal{B}}$ for batch $\mathcal{B}$
        \If{the batch satisfies the dark-field condition}
            \State $U_{\mathrm{pred}} \leftarrow U_{\mathrm{filt}}$
        \Else
            \State $U_{\mathrm{pred}} \leftarrow U_{\mathrm{filt}} + U_i^{\mathcal{B}}(\boldsymbol{r}_{\perp})$
        \EndIf
        \State $F_{\mathrm{pred}} \leftarrow \nabla_{\perp}(|U_{\mathrm{pred}}|^{2\gamma})$
        \State $\mathcal{L}_{\mathrm{FD-FPT}} \leftarrow \|F_{\mathrm{meas}}^{\mathcal{B}}-F_{\mathrm{pred}}\|_2^2 + \mathscr{R}(V_t)$
        \State Update $\tilde{V}_{t+1}$ using the gradient of $\mathcal{L}_{\mathrm{FD-FPT}}$ and set $t \leftarrow t+1$
    \EndFor
\EndFor
\State $V_{\mathrm{final}}(\boldsymbol{r}) \leftarrow \mathcal{F}_{3\mathrm{D}}^{-1}\{\tilde{V}_t(\boldsymbol{k})\}$
\State $n(\boldsymbol{r}) \leftarrow \sqrt{n_0^2+V_{\mathrm{final}}(\boldsymbol{r})/k_0^2}$
\end{algorithmic}
\endgroup
\end{algorithm*}

\section{Results}
Hereafter, spatial-domain FPT (SD-FPT) denotes the conventional intensity-domain baseline, whereas FD-FPT denotes the proposed feature-domain method. For conciseness, ``BF'' denotes the BF-only configuration, which uses only bright-field measurements, whereas ``DF'' denotes the BF+DF (dark-field-inclusive) configuration, which combines bright-field and dark-field measurements.

\subsection{Numerical validation using a lymph-node vascular phantom}

\begin{figure*}[!t]
    \centering
    \includegraphics[width=\linewidth]{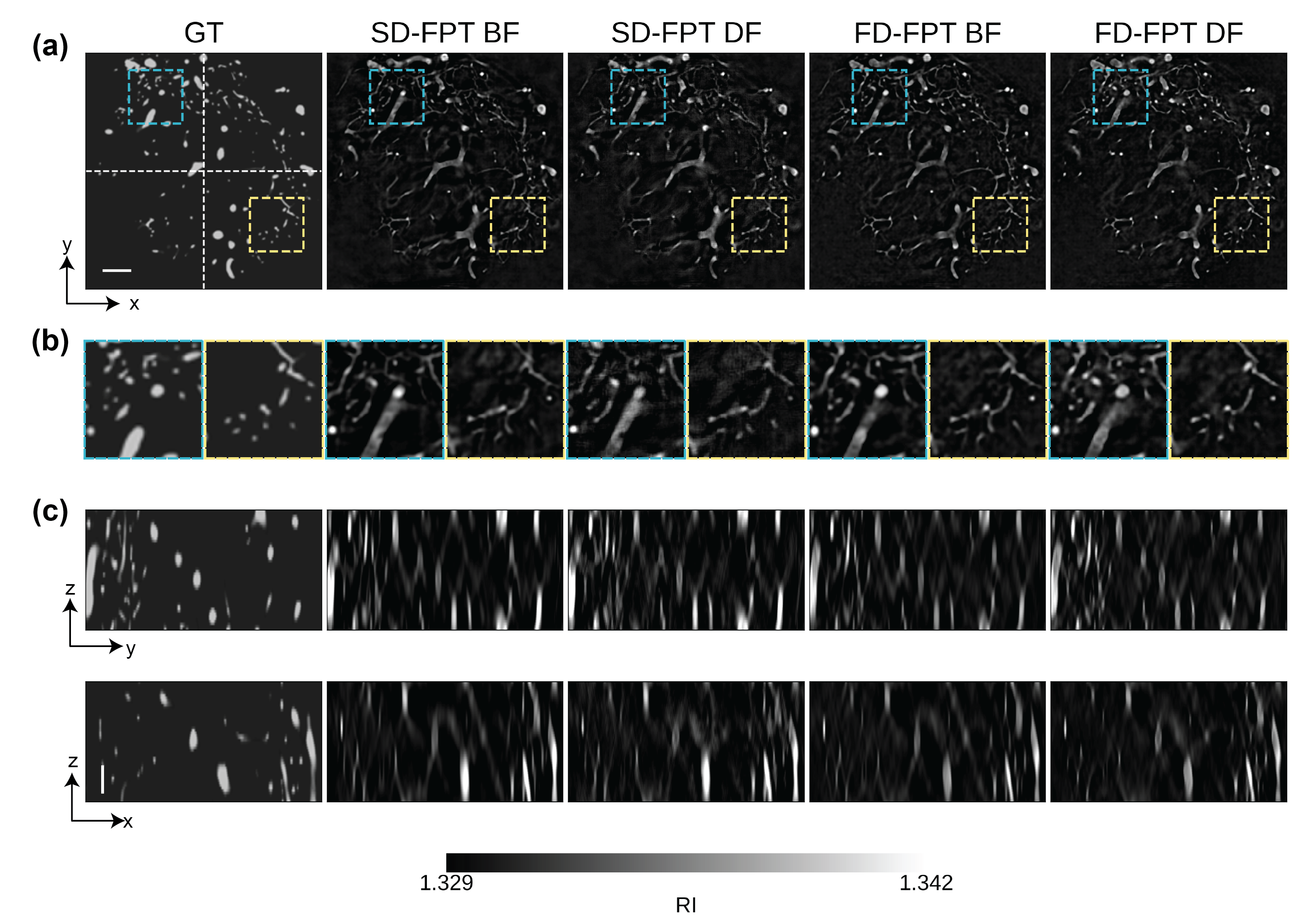}
    \caption{Simulation results for the lymph-node vascular phantom.
    (a) Ground-truth (GT) transverse RI slice and the corresponding
    SD-FPT and FD-FPT reconstructions ($z=0\,\mu\mathrm{m}$) under the
    BF and DF configurations.
    (b) Matched magnified views ($z=0\,\mu\mathrm{m}$) of the regions indicated
    by the cyan and yellow dashed boxes in (a). (c) Corresponding
    $y$--$z$ and $x$--$z$ orthogonal sections at the positions marked
    by the horizontal and vertical white dashed lines in the GT panel,
    respectively. In the panel labels, BF denotes the bright-field-only
    configuration, whereas DF denotes the dark-field-inclusive configuration. All images use the
    same RI display range of 1.329--1.342. Scale bar:
    $20\,\mu\mathrm{m}$.}
    \label{fig:results-sim}
\end{figure*}

To evaluate the feature-domain loss under controlled ground-truth
conditions, we used a $42\,\mu\mathrm{m}$-thick lymph-node
vascular phantom with complex vessel structures (RI = 1.34) embedded in a homogeneous background medium 
($n_0=1.33$), following Ref.~\cite{dong2025analytic}. Based on the
first Born approximation and an objective NA of 0.25, we generated
97 bright-field and 128 dark-field intensity measurements, with maximum illumination
NAs of 0.25 and 0.378, respectively. For this simulation, the BF
configuration used the 97 bright-field measurements alone, whereas the DF
configuration combined the 97 bright-field and 128 dark-field measurements
(225 measurements in total). Each configuration
was evaluated over 20 independent trials.

As shown in Figure~\ref{fig:results-sim}(a,b), SD-FPT recovered the
major vessels but exhibited stronger background modulation and
discontinuities in weak branches, particularly when the dark-field measurements
were included. In comparison, FD-FPT produced cleaner transverse
slices and preserved small vessels more continuously, with FD-FPT DF showing the closest agreement with the GT. The orthogonal
sections in Figure~\ref{fig:results-sim}(c) further indicate that
FD-FPT reduced axial elongation, with a larger improvement
in the DF configuration.

Across the 20 trials, FD-FPT increased the mean central-slice PSNR
by $0.72~\mathrm{dB}$ and $1.65~\mathrm{dB}$ relative to SD-FPT for
the BF and DF configurations, respectively. FD-FPT also retained higher
2D and 3D PSNR than the corresponding SD-FPT baselines, with FD-FPT DF
yielding the highest values (Supplementary Figure~S2).

\subsection{Experimental validation of spatial resolution}

The system consisted of an Olympus CKX41 microscope, a 201-element LED
array arranged on a circularly truncated $15\times15$ Cartesian grid
(4-mm pitch and 632-nm wavelength), a PCO.edge 5.5 sCMOS camera, and
a $10\times/0.25$ objective
[Figure~\ref{fig:fd-fpt-pipeline}(a)]. The LED array was positioned
68.5~mm from the sample, providing a maximum illumination NA of
0.378 and a synthetic NA of 0.628. According to
$\delta=\lambda/\mathrm{NA}_{\mathrm{syn}}$, the corresponding
predicted minimum resolvable spatial period was approximately
$1.01\,\mu\mathrm{m}$. Of the 201 intensity measurements, 97 were
bright-field measurements and 104 were dark-field measurements. The BF
configuration used the 97 bright-field measurements alone, whereas the DF
configuration combined all 201 measurements. All reconstructions were performed under the
first Born approximation.

\begin{figure*}[!t]
    \centering
    \includegraphics[width=\linewidth]{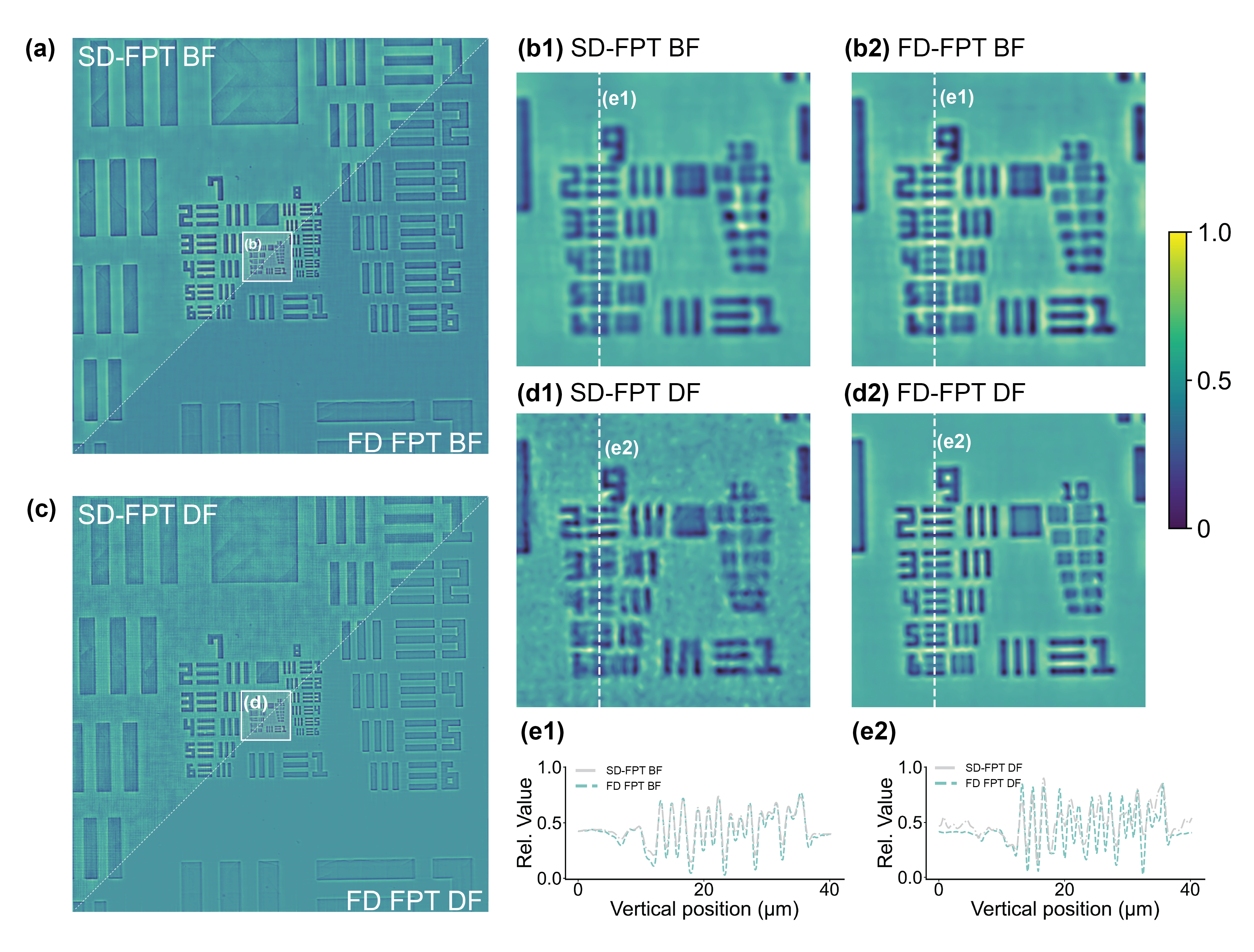}
    \caption{Experimental resolution characterization using a
    custom-fabricated phase-type USAF 1951 target. (a,c) Full-FOV
    maps ($z=0\,\mu\mathrm{m}$) of the normalized real part of the scattering potential
    reconstructed under the BF and DF configurations, respectively.
    Each map is diagonally divided between the registered SD-FPT
    (upper left) and FD-FPT (lower right) results. (b1,b2) Magnified
    BF reconstructions of the boxed region in (a) obtained using
    SD-FPT and FD-FPT, respectively. (d1,d2) Corresponding DF
    reconstructions of the boxed region in (c). (e1,e2) Normalized
    vertical profiles extracted along the matched dashed lines in
    the BF and DF reconstructions, respectively. Gray and cyan curves
    represent SD-FPT and FD-FPT. Each reconstruction map was
    independently normalized to $[0,1]$ and displayed using the same
    normalized color range.}
    \label{fig:usaf-resolution}
\end{figure*}

We evaluated the lateral resolution using a phase-type USAF 1951
metasurface target consisting of 630-nm-tall SiN structures on a fused-silica
substrate (Figure~\ref{fig:usaf-resolution}). Under the BF
configuration, both methods recovered the bar patterns, with FD-FPT
showing a more spatially uniform background and clearer edges. Under
the DF configuration, SD-FPT exhibited strong background fluctuations
that obscured the finest elements, whereas FD-FPT reduced these
artifacts and made Group 9 Element 6 distinguishable. This
element has a line-pair period of approximately $1.10\,\mu\mathrm{m}$
and a line width of approximately $0.55\,\mu\mathrm{m}$, consistent
with the theoretical minimum spatial period. The profiles in
Figure~\ref{fig:usaf-resolution}(e1,e2) show similar structural
artifacts under the BF configuration but a clearer peak-to-valley
separation for FD-FPT under the DF configuration.

\subsection{Quantitative validation using microspheres}

\begin{figure*}[!t]
    \centering
    \includegraphics[width=\linewidth]{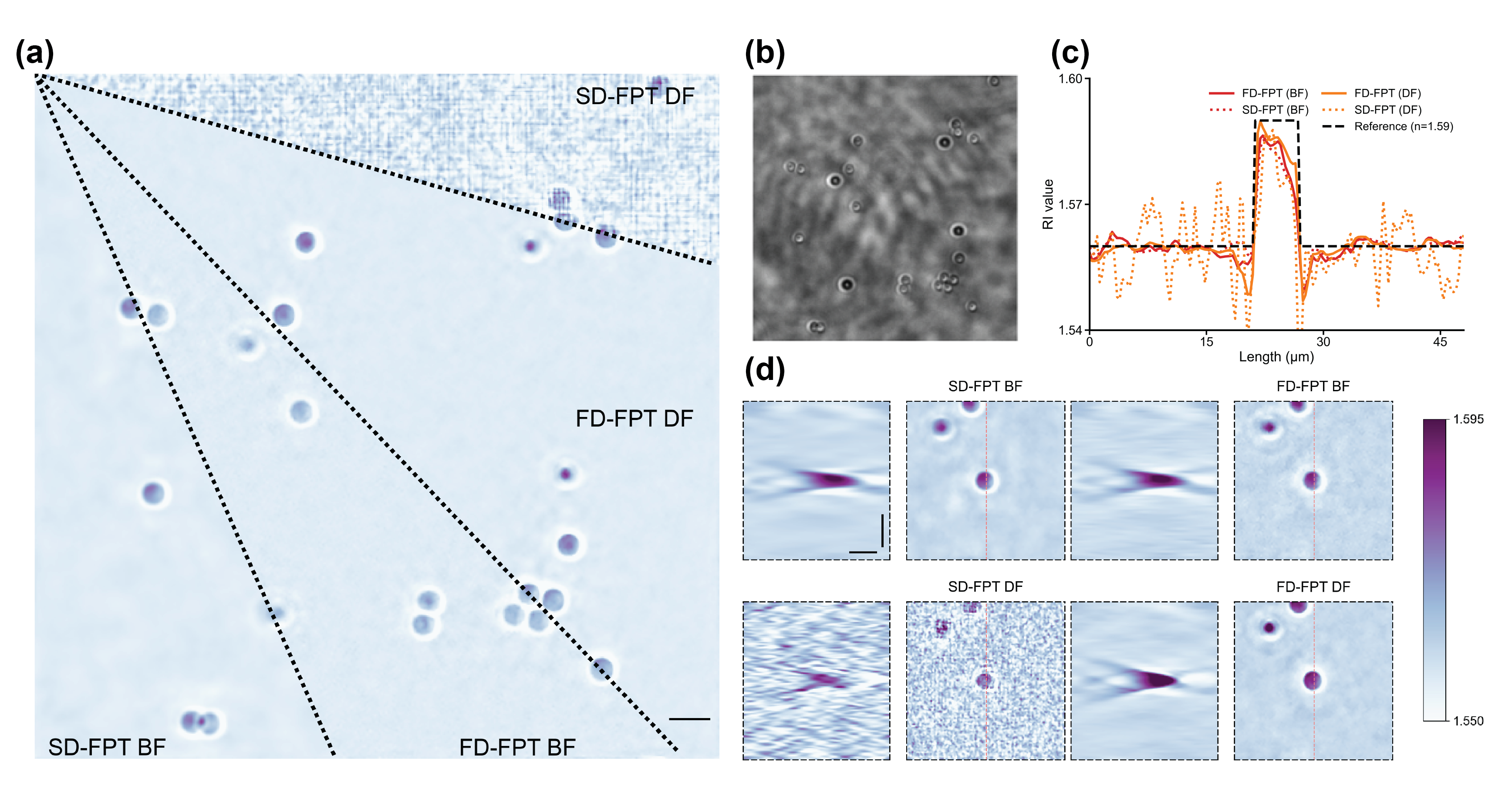}
    \caption{Experimental RI comparison using 6-$\mu\mathrm{m}$
    polystyrene microspheres embedded in NOA81. (a) Registered
    full-field transverse RI maps ($z=0\,\mu\mathrm{m}$) reconstructed using SD-FPT and
    FD-FPT under the BF and DF configurations. The dotted
    diagonal lines separate the four reconstructions, and the dashed
    box indicates the region analyzed in (d). (b) Representative raw
    intensity measurement. (c) RI profiles extracted along the red
    dashed lines in (d); the black dashed curve denotes the nominal
    profile based on $n=1.59$ for polystyrene and $n=1.56$ for NOA81.
    (d) Matched $y$--$z$ axial sections (left) and $x$--$y$ transverse
    sections (right; $z=0\,\mu\mathrm{m}$) of the selected microsphere. All RI maps use the
    same display range of 1.550--1.595. Scale bar in (a):
    $10\,\mu\mathrm{m}$. The horizontal and vertical scale bars in
    (d) indicate $10\,\mu\mathrm{m}$ along $z$ and $y$, respectively.}
    \label{fig:results-beads}
\end{figure*}

To evaluate RI recovery and robustness when incorporating dark-field
measurements, we imaged 6-$\mu\mathrm{m}$ polystyrene microspheres
embedded in NOA81 optical resin (Figure~\ref{fig:results-beads}). Nominal RI values
of 1.59 and 1.56 were used as references for polystyrene and NOA81,
respectively. As shown in Figure~\ref{fig:results-beads}(a), SD-FPT DF
exhibits pronounced grid-like background fluctuations that partially
obscure the microspheres. In comparison, FD-FPT DF reduces
these artifacts and preserves more clearly defined
microsphere boundaries.

The profiles in Figure~\ref{fig:results-beads}(c) show that both SD-FPT and FD-FPT methods yield comparable background stability and microsphere RI
under the BF configuration. Under the DF configuration,
however, the SD-FPT estimate fluctuates strongly around the nominal
medium RI, whereas FD-FPT maintains a stable background near
$n=1.56$ and recovers an elevated microsphere RI approaching the
nominal value of $n=1.59$. The matched sections in
Figure~\ref{fig:results-beads}(d) show a similar trend. The BF
reconstructions exhibit only minor differences, whereas FD-FPT DF
reduces the structured background artifacts and
produces a cleaner, more compact axial distribution than SD-FPT DF.
Although residual axial elongation remains because of the
missing-cone effect \cite{lim2015comparative}, these results indicate
that the feature-domain loss primarily improves robustness to the
inclusion of dark-field measurements rather than substantially altering the
BF reconstruction.

\subsection{Three-dimensional imaging of an Oedogonium specimen}
\begin{figure*}[!t]
    \centering
    \includegraphics[width=\linewidth]{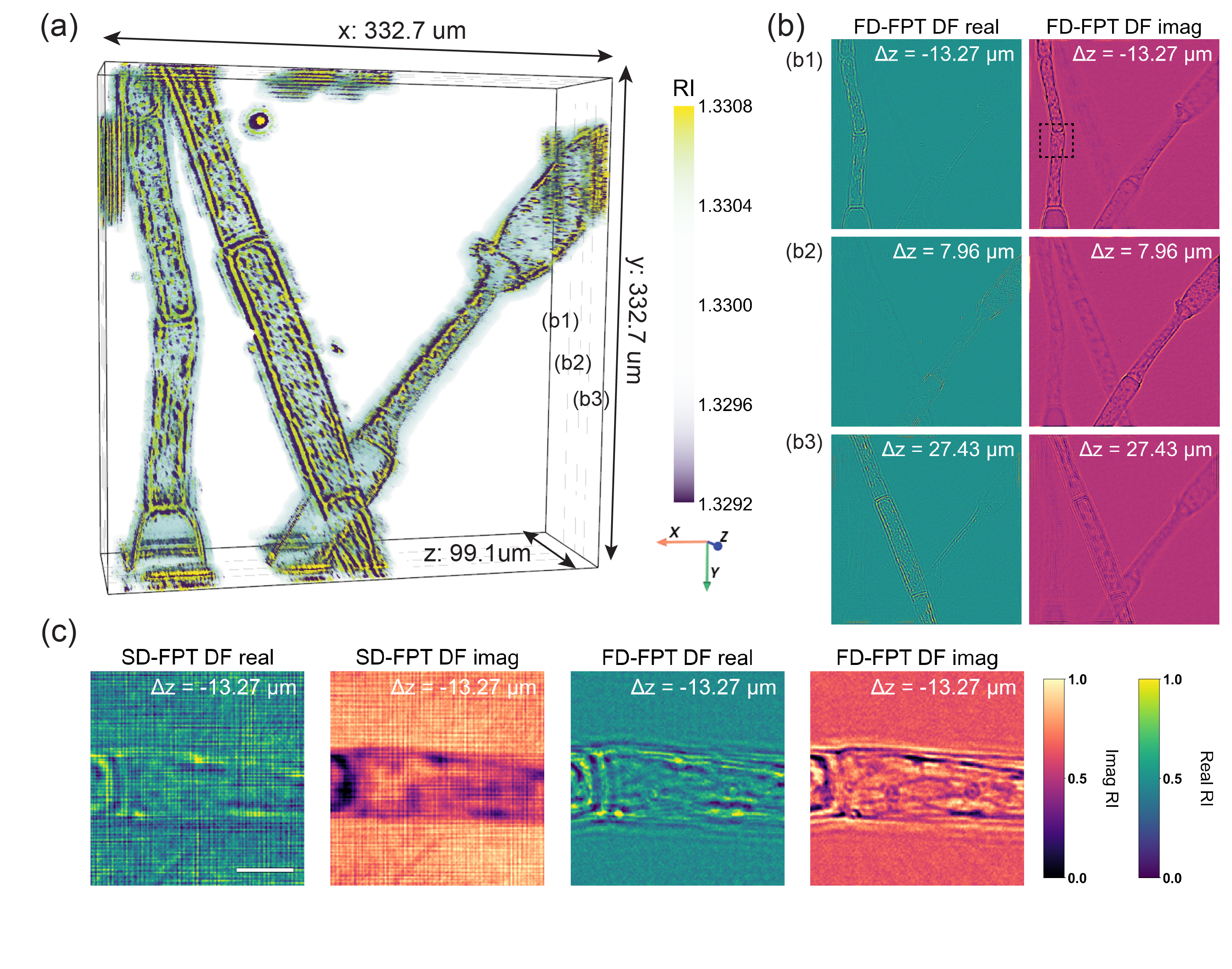}
    \caption{Volumetric reconstruction of a whole-mount
    \textit{Oedogonium} specimen. (a) Three-dimensional rendering of
    the RI distribution reconstructed using FD-FPT under the DF
    configuration. The color scale spans RI values from 1.3292 to
    1.3308. (b) Real and imaginary components reconstructed using
    FD-FPT DF at $\Delta z=-13.27$, $+7.96$, and
    $+27.43\,\mu\mathrm{m}$. The dashed box in the imaginary
    component of (b1) indicates the region compared in (c).
    (c) Matched SD-FPT and FD-FPT reconstructions under the DF
    configuration at $\Delta z=-13.27\,\mu\mathrm{m}$. In (c), the
    real and imaginary components
    were independently normalized to $[0,1]$ for visualization.
    Scale bar in (c): $20\,\mu\mathrm{m}$.}
    \label{fig:results-oedogonium}
\end{figure*}
To evaluate volumetric recovery in a thick biological specimen with
complex internal morphology, we imaged a whole-mount
\textit{Oedogonium} specimen. The reconstructed volume comprised
$881\times881\times113$ voxels with transverse and axial sampling
intervals of $0.378\,\mu\mathrm{m}$ and $0.885\,\mu\mathrm{m}$,
respectively, corresponding to a physical volume of approximately
$332.7\times332.7\times99.1\,\mu\mathrm{m}^3$. This
87.7-megavoxel volume captures several overlapping, unbranched
filaments distributed at different depths, as shown in
Figure~\ref{fig:results-oedogonium}(a).

The complex-valued reconstruction provides complementary structural
contrasts. The real component primarily reflects RI variations and
therefore delineates phase-related morphology, including the cell
walls, transverse septa, and intracellular boundaries. The imaginary
component reflects optical attenuation associated with absorption
and unresolved scattering. The sections at $\Delta z=-13.27$,
$+7.96$, and $+27.43\,\mu\mathrm{m}$ in
Figure~\ref{fig:results-oedogonium}(b) span an axial interval of
$40.70\,\mu\mathrm{m}$ and reveal different filament segments and
intracellular features throughout the reconstructed volume, illustrating
the depth-dependent morphology of the specimen.

The enlarged region in Figure~\ref{fig:results-oedogonium}(c) contains
continuous, interconnected intracellular bands oriented predominantly
along the cell axis. This organization is morphologically consistent
with the reticulate, parietal chloroplast characteristic of
\textit{Oedogonium} \cite{xiong2022morphology}.
SD-FPT DF produces pronounced grid-like background artifacts that
partially obscure the cell boundary and intracellular organization.
In comparison, FD-FPT DF suppresses these artifacts and more clearly
preserves the longitudinal cell walls, transverse septum, and
chloroplast-like network in both the real and imaginary components.
These comparisons indicate that the feature-domain loss improves the
visibility of sample structures and reduces reconstruction artifacts
when using the DF configuration.

\subsection{Large-FOV volumetric imaging of a mouse adrenal gland tissue slide}
\enlargethispage{\baselineskip}

\begin{figure*}[!t]
    \centering
    \includegraphics[width=0.95\linewidth]{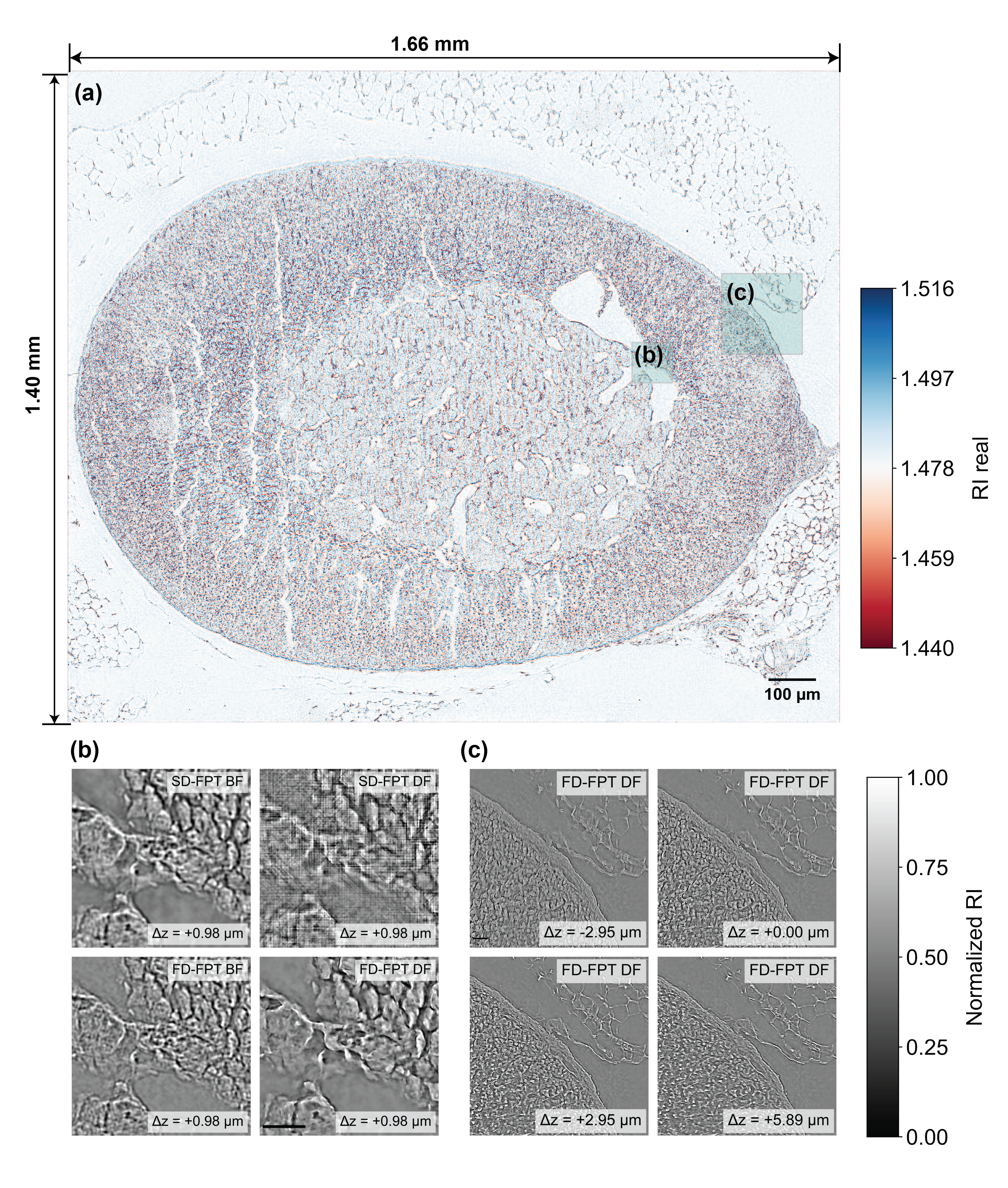}
    \caption{Gigavoxel-scale volumetric reconstruction of a mouse
    adrenal gland section. The reconstructed volume contains
    $4236\times5020\times85$ voxels, corresponding to approximately
    1.81 gigavoxels and a physical volume of
    $1664\times1404\times83.5\,\mu\mathrm{m}^3$.
    (a) Full-FOV ($1.664\times1.404\,\mathrm{mm}^2$) map of the real
    RI component ($z=0\,\mu\mathrm{m}$). The shaded boxes indicate the regions examined in
    (b) and (c), respectively.
    (b) Matched transverse reconstructions at
    $\Delta z=+0.98\,\mu\mathrm{m}$ obtained using SD-FPT and FD-FPT
    under the BF and DF configurations. (c) Depth-resolved transverse
    reconstructions of the capsular and periadrenal region obtained
    using FD-FPT DF at $\Delta z=-2.95$, $0$, $+2.95$, and
    $+5.89\,\mu\mathrm{m}$.
    The color scale in (a) spans reconstructed real-RI values from
    1.440 to 1.516, whereas the grayscale bar represents the
    normalized display range in (b,c). Scale bars in (a):
    $100\,\mu\mathrm{m}$, in (b,c): $20\,\mu\mathrm{m}$.}
    \label{fig:results-wide_adrenal}
\end{figure*}

To evaluate scalability to gigavoxel imaging, we applied FD-FPT to a
millimeter-scale mouse adrenal gland section
(Figure~\ref{fig:results-wide_adrenal}). The resulting 1.81-gigavoxel
dataset combines
millimeter-scale coverage with depth-resolved volumetric sampling
rather than representing a single-plane wide-field reconstruction.

At the tissue scale, the RI distribution in
Figure~\ref{fig:results-wide_adrenal}(a) reveals a peripheral band of
densely textured glandular tissue surrounding a morphologically
distinct central region. This organization is consistent with the
adrenal cortex and medulla, respectively, while the thin outer
boundary and surrounding polygonal structures are consistent with
the fibrous capsule and periadrenal adipose tissue
\cite{nicolaides2023adrenal}. Variations in RI contrast within
the gland likely reflect differences in cellular packing, lipid
content, stromal organization, and vascular spaces. Because these
assignments are based on label-free morphology, individual cortical
zones and specific cell types cannot be identified conclusively
without histological or molecular markers.

The matched reconstructions in
Figure~\ref{fig:results-wide_adrenal}(b) show that FD-FPT preserves
sharper cellular and stromal boundaries than SD-FPT under the BF
configuration. Under the DF configuration, SD-FPT
exhibits pronounced structured background artifacts that partially
obscure the local tissue organization, whereas FD-FPT suppresses
these artifacts and recovers clearer cellular outlines and
intercellular spaces.

The depth series in Figure~\ref{fig:results-wide_adrenal}(c) spans
$8.84\,\mu\mathrm{m}$ across the gland boundary. The densely textured
glandular region, thin capsular interface, and polygonal outlines of
the adjacent adipose tissue change progressively with depth,
consistent with depth-dependent localization of distinct tissue
compartments across the large FOV. Together, these results show that
FD-FPT supports gigavoxel-scale volumetric reconstruction with
cellular-scale structural contrast, with the largest
improvement observed under the DF configuration.

\section{Discussion and Conclusion}

In this study, we introduced feature-domain Fourier ptychographic
tomography (FD-FPT) to improve the robustness of three-dimensional RI
reconstruction, particularly when low-SNR dark-field measurements are
included. By evaluating data fidelity after transverse-gradient
extraction, FD-FPT is less sensitive to slowly varying intensity
mismatches while retaining edge information. Compared with
SD-FPT, it produced cleaner backgrounds, more continuous fine
structures, and reduced axial spreading, with the largest improvement
observed under the DF configuration. FD-FPT recovered structures near
the synthetic-aperture cutoff, including USAF Group 9 Element 6 with
a $1.10\,\mu\mathrm{m}$ line-pair period, stabilized the quantitative
RI reconstruction of microspheres, made chloroplast-like structures
in \textit{Oedogonium} more clearly visible, and enabled a 1.81-gigavoxel
reconstruction of mouse adrenal tissue. These results indicate that
feature-domain optimization improves robustness when incorporating
high-angle dark-field measurements without changing the physical bandwidth
of the imaging system.

The main limitations arise from the forward model and frequency coverage. The first Born approximation may become inaccurate for thick or strongly scattering specimens, while incomplete axial-frequency coverage continues to produce residual missing-cone artifacts. In addition, the fixed first-order gradient used to extract features for the feature-domain loss may suppress sensitivity to low-frequency information in exchange for resolving high-frequency biological features with better sensitivity and robustness.

Future work could combine feature-domain optimization with
multiple-scattering models and improved axial-frequency sampling.
Adaptive features, including higher-order gradients
\cite{xu2025imaging}, wavelets \cite{wu2025wavelet}, or learned
representations \cite{zeiler2014visualizing,simonyan2014very}, may
further improve robustness across different specimens and noise
conditions. Memory-efficient neural fields
\cite{liu2022recovery,zhang2025whole} or Gaussian splatting
\cite{zhang2025gaussian}, together with compact imaging hardware
\cite{aidukas2019low,lee2021smartphone}, could advance FD-FPT toward
scalable, high-resolution, label-free three-dimensional imaging for
biomedical research.

\begin{backmatter}

\bmsection{Disclosures}
The authors declare no conflict of interest.

\bmsection{Data Availability Statement}
The data that support the findings of this study are available from the corresponding author upon reasonable request.

\bmsection{Supplemental Document}
Additional supporting information is provided in the accompanying Supporting Information document.

\end{backmatter}

\bibliography{FD_FPT_arXiv}

\begin{thebibliography}{10}
\newcommand{\enquote}[1]{``#1''}

\bibitem{hawkes2019springer}
P.~W. Hawkes and J.~C. Spence, \emph{Springer handbook of microscopy} (Springer
  Nature, 2019).

\bibitem{mertz2019strategies}
J.~Mertz, \enquote{Strategies for volumetric imaging with a fluorescence
  microscope,} {\protect\JournalTitle{Optica}} \textbf{6}, 1261--1268 (2019).

\bibitem{lee2013quantitative}
K.~Lee, K.~Kim, J.~Jung, J.~Heo, S.~Cho, S.~Lee, G.~Chang, Y.~Jo, H.~Park, and
  Y.~Park, \enquote{Quantitative phase imaging techniques for the study of cell
  pathophysiology: from principles to applications,}
  {\protect\JournalTitle{Sensors}} \textbf{13}, 4170--4191 (2013).

\bibitem{wolf1969three}
E.~Wolf, \enquote{Three-dimensional structure determination of semi-transparent
  objects from holographic data,} {\protect\JournalTitle{Optics
  communications}} \textbf{1}, 153--156 (1969).

\bibitem{devaney2012mathematical}
A.~J. Devaney, \emph{Mathematical foundations of imaging, tomography and
  wavefield inversion} (Cambridge University Press, 2012).

\bibitem{lauer2002new}
V.~Lauer, \enquote{New approach to optical diffraction tomography yielding a
  vector equation of diffraction tomography and a novel tomographic
  microscope,} {\protect\JournalTitle{Journal of microscopy}} \textbf{205},
  165--176 (2002).

\bibitem{sung2009optical}
Y.~Sung, W.~Choi, C.~Fang-Yen, K.~Badizadegan, R.~R. Dasari, and M.~S. Feld,
  \enquote{Optical diffraction tomography for high resolution live cell
  imaging,} {\protect\JournalTitle{Optics express}} \textbf{17}, 266--277
  (2009).

\bibitem{bhaduri2014diffraction}
B.~Bhaduri, C.~Edwards, H.~Pham, R.~Zhou, T.~H. Nguyen, L.~L. Goddard, and
  G.~Popescu, \enquote{Diffraction phase microscopy: principles and
  applications in materials and life sciences,} {\protect\JournalTitle{Advances
  in Optics and Photonics}} \textbf{6}, 57--119 (2014).

\bibitem{jin2017tomographic}
D.~Jin, R.~Zhou, Z.~Yaqoob, and P.~T. So, \enquote{Tomographic phase
  microscopy: principles and applications in bioimaging,}
  {\protect\JournalTitle{Journal of the Optical Society of America B}}
  \textbf{34}, B64--B77 (2017).

\bibitem{kamilov2015learning}
U.~S. Kamilov, I.~N. Papadopoulos, M.~H. Shoreh, A.~Goy, C.~Vonesch, M.~Unser,
  and D.~Psaltis, \enquote{Learning approach to optical tomography,}
  {\protect\JournalTitle{Optica}} \textbf{2}, 517--522 (2015).

\bibitem{maleki1993phase}
M.~H. Maleki and A.~J. Devaney, \enquote{Phase-retrieval and intensity-only
  reconstruction algorithms for optical diffraction tomography,}
  {\protect\JournalTitle{Journal of the Optical Society of America A}}
  \textbf{10}, 1086--1092 (1993).

\bibitem{gbur2004information}
G.~Gbur and E.~Wolf, \enquote{The information content of the scattered
  intensity in diffraction tomography,} {\protect\JournalTitle{Information
  Sciences}} \textbf{162}, 3--20 (2004).

\bibitem{horstmeyer2016diffraction}
R.~Horstmeyer, J.~Chung, X.~Ou, G.~Zheng, and C.~Yang, \enquote{Diffraction
  tomography with fourier ptychography,} {\protect\JournalTitle{Optica}}
  \textbf{3}, 827--835 (2016).

\bibitem{li2017optical}
J.~Li, Q.~Chen, J.~Zhang, Z.~Zhang, Y.~Zhang, and C.~Zuo, \enquote{Optical
  diffraction tomography microscopy with transport of intensity equation using
  a light-emitting diode array,} {\protect\JournalTitle{Optics and lasers in
  engineering}} \textbf{95}, 26--34 (2017).

\bibitem{ling2018high}
R.~Ling, W.~Tahir, H.-Y. Lin, H.~Lee, and L.~Tian, \enquote{High-throughput
  intensity diffraction tomography with a computational microscope,}
  {\protect\JournalTitle{Biomedical optics express}} \textbf{9}, 2130--2141
  (2018).

\bibitem{zuo2020wide}
C.~Zuo, J.~Sun, J.~Li, A.~Asundi, and Q.~Chen, \enquote{Wide-field
  high-resolution 3d microscopy with fourier ptychographic diffraction
  tomography,} {\protect\JournalTitle{Optics and Lasers in Engineering}}
  \textbf{128}, 106003 (2020).

\bibitem{baek2021intensity}
Y.~Baek and Y.~Park, \enquote{Intensity-based holographic imaging via
  space-domain kramers--kronig relations,} {\protect\JournalTitle{Nature
  Photonics}} \textbf{15}, 354--360 (2021).

\bibitem{li2022transport}
J.~Li, N.~Zhou, J.~Sun, S.~Zhou, Z.~Bai, L.~Lu, Q.~Chen, and C.~Zuo,
  \enquote{Transport of intensity diffraction tomography with
  non-interferometric synthetic aperture for three-dimensional label-free
  microscopy,} {\protect\JournalTitle{Light: Science \& Applications}}
  \textbf{11}, 154 (2022).

\bibitem{zhou2022transport}
S.~Zhou, J.~Li, J.~Sun, N.~Zhou, H.~Ullah, Z.~Bai, Q.~Chen, and C.~Zuo,
  \enquote{Transport-of-intensity fourier ptychographic diffraction tomography:
  defying the matched illumination condition,} {\protect\JournalTitle{Optica}}
  \textbf{9}, 1362--1373 (2022).

\bibitem{sun2022high}
M.~Sun, L.~Shao, J.~Zhang, Y.~Zhu, P.~Wu, Y.~Wang, Z.~Diao, Q.~Mu, D.~Li,
  H.~Wang \emph{et~al.}, \enquote{High-resolution 3d fourier ptychographic
  reconstruction using a hemispherical illumination source with
  multiplexed-coded strategy,} {\protect\JournalTitle{Biomedical Optics
  Express}} \textbf{13}, 2050--2067 (2022).

\bibitem{zhou2020diffraction}
K.~C. Zhou and R.~Horstmeyer, \enquote{Diffraction tomography with a deep image
  prior,} {\protect\JournalTitle{Optics express}} \textbf{28}, 12872--12896
  (2020).

\bibitem{zheng2013wide}
G.~Zheng, R.~Horstmeyer, and C.~Yang, \enquote{Wide-field, high-resolution
  fourier ptychographic microscopy,} {\protect\JournalTitle{Nature photonics}}
  \textbf{7}, 739--745 (2013).

\bibitem{zheng2021concept}
G.~Zheng, C.~Shen, S.~Jiang, P.~Song, and C.~Yang, \enquote{Concept,
  implementations and applications of fourier ptychography,}
  {\protect\JournalTitle{Nature Reviews Physics}} \textbf{3}, 207--223 (2021).

\bibitem{jiang2023spatial}
S.~Jiang, P.~Song, T.~Wang, L.~Yang, R.~Wang, C.~Guo, B.~Feng, A.~Maiden, and
  G.~Zheng, \enquote{Spatial-and fourier-domain ptychography for
  high-throughput bio-imaging,} {\protect\JournalTitle{Nature protocols}}
  \textbf{18}, 2051--2083 (2023).

\bibitem{xu2024fourier}
F.~Xu, Z.~Wu, C.~Tan, Y.~Liao, Z.~Wang, K.~Chen, and A.~Pan, \enquote{Fourier
  ptychographic microscopy 10 years on: a review,}
  {\protect\JournalTitle{Cells}} \textbf{13}, 324 (2024).

\bibitem{tian20153d}
L.~Tian and L.~Waller, \enquote{3d intensity and phase imaging from light field
  measurements in an led array microscope,} {\protect\JournalTitle{optica}}
  \textbf{2}, 104--111 (2015).

\bibitem{li2018multi}
P.~Li and A.~Maiden, \enquote{Multi-slice ptychographic tomography,}
  {\protect\JournalTitle{Scientific reports}} \textbf{8}, 2049 (2018).

\bibitem{chowdhury2019high}
S.~Chowdhury, M.~Chen, R.~Eckert, D.~Ren, F.~Wu, N.~Repina, and L.~Waller,
  \enquote{High-resolution 3d refractive index microscopy of
  multiple-scattering samples from intensity images,}
  {\protect\JournalTitle{Optica}} \textbf{6}, 1211--1219 (2019).

\bibitem{lee2022inverse}
M.~Lee, H.~Hugonnet, and Y.~Park, \enquote{Inverse problem solver for multiple
  light scattering using modified born series,} {\protect\JournalTitle{Optica}}
  \textbf{9}, 177--182 (2022).

\bibitem{lim2015comparative}
J.~Lim, K.~Lee, K.~H. Jin, S.~Shin, S.~Lee, Y.~Park, and J.~C. Ye,
  \enquote{Comparative study of iterative reconstruction algorithms for missing
  cone problems in optical diffraction tomography,}
  {\protect\JournalTitle{Optics express}} \textbf{23}, 16933--16948 (2015).

\bibitem{dong2025analytic}
Z.~Dong, H.~Zhou, R.~Cao, O.~Zhang, S.~Zhao, P.~Lyu, R.~Alcalde, and C.~Yang,
  \enquote{Analytic fourier ptychotomography for aberration-free and
  high-resolution volumetric refractive index imaging,}
  {\protect\JournalTitle{Nature Communications}}  (2025).

\bibitem{yeh2015experimental}
L.-H. Yeh, J.~Dong, J.~Zhong, L.~Tian, M.~Chen, G.~Tang, M.~Soltanolkotabi, and
  L.~Waller, \enquote{Experimental robustness of fourier ptychography phase
  retrieval algorithms,} {\protect\JournalTitle{Optics express}} \textbf{23},
  33214--33240 (2015).

\bibitem{pan2017system}
A.~Pan, Y.~Zhang, T.~Zhao, Z.~Wang, D.~Dan, M.~Lei, and B.~Yao, \enquote{System
  calibration method for fourier ptychographic microscopy,}
  {\protect\JournalTitle{Journal of biomedical optics}} \textbf{22},
  096005--096005 (2017).

\bibitem{zhang2023elfpie}
S.~Zhang, T.~T. Berendschot, and J.~Zhou, \enquote{Elfpie: an error-laxity
  fourier ptychographic iterative engine,} {\protect\JournalTitle{Signal
  Processing}} \textbf{210}, 109088 (2023).

\bibitem{chen2025uncertainty}
N.~Chen, Y.~Wu, C.~Tan, L.~Cao, J.~Wang, and E.~Y. Lam,
  \enquote{Uncertainty-aware fourier ptychography,}
  {\protect\JournalTitle{Light: Science \& Applications}} \textbf{14}, 236
  (2025).

\bibitem{zhang2025whole}
S.~Zhang and L.~Cao, \enquote{Whole-field, high-resolution fourier ptychography
  with neural pupil engineering,} {\protect\JournalTitle{Optica}} \textbf{12},
  1615--1624 (2025).

\bibitem{zhao2024deep}
Q.~Zhao, R.~Wang, S.~Zhang, T.~Wang, P.~Song, and G.~Zheng,
  \enquote{Deep-ultraviolet fourier ptychography (duv-fp) for label-free
  biochemical imaging via feature-domain optimization,}
  {\protect\JournalTitle{APL photonics}} \textbf{9} (2024).

\bibitem{zhang2025high}
S.~Zhang, A.~Pan, H.~Sun, Y.~Tan, and L.~Cao, \enquote{High-fidelity
  computational microscopy via feature-domain phase retrieval,}
  {\protect\JournalTitle{Advanced Science}} \textbf{12}, 2413975 (2025).

\bibitem{wu2025wavelet}
H.~Wu, J.~Wang, H.~Pan, J.~Lyu, S.~Zhang, and J.~Zhou, \enquote{Wavelet-forward
  family enabling stitching-free full-field fourier ptychographic microscopy,}
  {\protect\JournalTitle{Laser \& Photonics Reviews}} \textbf{19}, 2401183
  (2025).

\bibitem{baydin2018automatic}
A.~G. Baydin, B.~A. Pearlmutter, A.~A. Radul, and J.~M. Siskind,
  \enquote{Automatic differentiation in machine learning: a survey,}
  {\protect\JournalTitle{Journal of machine learning research}} \textbf{18},
  1--43 (2018).

\bibitem{paszke2017automatic}
A.~Paszke, S.~Gross, S.~Chintala, G.~Chanan, E.~Yang, Z.~DeVito, Z.~Lin,
  A.~Desmaison, L.~Antiga, and A.~Lerer, \enquote{Automatic differentiation in
  pytorch,} {\protect\JournalTitle{NIPS 2017 Autodiff Workshop}}  (2017).

\bibitem{park2018quantitative}
Y.~Park, C.~Depeursinge, and G.~Popescu, \enquote{Quantitative phase imaging in
  biomedicine,} {\protect\JournalTitle{Nature photonics}} \textbf{12}, 578--589
  (2018).

\bibitem{nguyen2022quantitative}
T.~L. Nguyen, S.~Pradeep, R.~L. Judson-Torres, J.~Reed, M.~A. Teitell, and
  T.~A. Zangle, \enquote{Quantitative phase imaging: recent advances and
  expanding potential in biomedicine,} {\protect\JournalTitle{ACS nano}}
  \textbf{16}, 11516--11544 (2022).

\bibitem{muller2015theory}
P.~M{\"u}ller, M.~Sch{\"u}rmann, and J.~Guck, \enquote{The theory of
  diffraction tomography,} {\protect\JournalTitle{arXiv preprint
  arXiv:1507.00466}}  (2015).

\bibitem{kak2001principles}
A.~C. Kak and M.~Slaney, \emph{Principles of computerized tomographic imaging}
  (SIAM, 2001).

\bibitem{xiong2022morphology}
Q.~Xiong, Y.~Chen, Q.~Dai, B.~Liu, and G.~Liu, \enquote{Morphology and
  molecular phylogeny of genus oedogonium (oedogoniales, chlorophyta) from
  china,} {\protect\JournalTitle{Plants}} \textbf{11}, 2422 (2022).

\bibitem{nicolaides2023adrenal}
N.~C. Nicolaides, H.~S. Willenberg, S.~R. Bornstein, and G.~P. Chrousos,
  \enquote{Adrenal cortex: Embryonic development, anatomy, histology and
  physiology,} in \emph{Endotext,}  K.~R. Feingold, R.~A. Adler, S.~F. Ahmed,
  B.~Anawalt, M.~R. Blackman, A.~Boyce, G.~P. Chrousos, E.~Corpas, W.~W.
  de~Herder, K.~Dhatariya, K.~Dungan, J.~Hofland, S.~Kalra, G.~Kaltsas,
  N.~Kapoor, C.~Koch, P.~A. Kopp, M.~Korbonits, C.~S. Kovacs, W.~Kuohung,
  B.~Laferrere, M.~Levy, E.~A. McGee, R.~McLachlan, M.~New, J.~Purnell,
  R.~Sahay, A.~S. Shah, F.~Singer, M.~A. Sperling, C.~A. Stratakis, D.~L.
  Trence, and D.~P. Wilson, eds. (MDText.com, Inc., South Dartmouth, MA, 2023).
  Updated June 12, 2023.

\bibitem{xu2025imaging}
J.~Xu, H.~Chen, Y.~Liao, T.~Feng, and A.~Pan, \enquote{Imaging through weak
  scattering media via high-order fourier ptychographic microscopy,}
  {\protect\JournalTitle{Optics \& Laser Technology}} \textbf{188}, 112890
  (2025).

\bibitem{zeiler2014visualizing}
M.~D. Zeiler and R.~Fergus, \enquote{Visualizing and understanding
  convolutional networks,} in \emph{European conference on computer vision,}
  (Springer, 2014), pp. 818--833.

\bibitem{simonyan2014very}
K.~Simonyan and A.~Zisserman, \enquote{Very deep convolutional networks for
  large-scale image recognition,} {\protect\JournalTitle{arXiv preprint
  arXiv:1409.1556}}  (2014).

\bibitem{liu2022recovery}
R.~Liu, Y.~Sun, J.~Zhu, L.~Tian, and U.~S. Kamilov, \enquote{Recovery of
  continuous 3d refractive index maps from discrete intensity-only measurements
  using neural fields,} {\protect\JournalTitle{Nature Machine Intelligence}}
  \textbf{4}, 781--791 (2022).

\bibitem{zhang2025gaussian}
S.~Zhang and L.~Cao, \enquote{Gaussian splatting holography,}
  {\protect\JournalTitle{arXiv preprint arXiv:2509.20774}}  (2025).

\bibitem{aidukas2019low}
T.~Aidukas, R.~Eckert, A.~R. Harvey, L.~Waller, and P.~C. Konda,
  \enquote{Low-cost, sub-micron resolution, wide-field computational microscopy
  using opensource hardware,} {\protect\JournalTitle{Scientific reports}}
  \textbf{9}, 7457 (2019).

\bibitem{lee2021smartphone}
K.~C. Lee, K.~Lee, J.~Jung, S.~H. Lee, D.~Kim, and S.~A. Lee, \enquote{A
  smartphone-based fourier ptychographic microscope using the display screen
  for illumination,} {\protect\JournalTitle{Acs Photonics}} \textbf{8},
  1307--1315 (2021).

\end{thebibliography}


\begin{thebibliography}{1}
\newcommand{\enquote}[1]{``#1''}

\bibitem{zhang2024fpm}
S.~Zhang, A.~Wang, J.~Xu, T.~Feng, J.~Zhou, and A.~Pan, \enquote{{FPM-WSI}:
  Fourier ptychographic whole slide imaging via feature-domain
  backdiffraction,} {\protect\JournalTitle{Optica}} \textbf{11}, 634--646
  (2024).

\bibitem{zhang2025high}
S.~Zhang, A.~Pan, H.~Sun, Y.~Tan, and L.~Cao, \enquote{High-fidelity
  computational microscopy via feature-domain phase retrieval,}
  {\protect\JournalTitle{Advanced Science}} \textbf{12}, 2413975 (2025).

\end{thebibliography}

\end{document}